\documentclass[a4paper]{jpconf}
\usepackage{graphicx}
\begin{document}
\title{Probing the symmetry energy with isospin ratio from nucleons to fragments}

\author{Yingxun Zhang$^1$, Zhuxia Li$^1$, Chengshuang Zhou$^{1,2}$, Jixian Chen$^{1,2}$, M.Colonna$^3$, P.Danielewicz$^4$ and M.B.Tsang$^4$}
\address{$^1$ China Institute of Atomic Energy, P.O. Box 275 (10), Beijing 102413, P.R. China}
\address{$^2$ College of Physics and Technology,Guangxi Normal University, Guilin 541004,P.R. China}
\address{$^3$ INFN, Laboratori Nazionali del Sud, Catania, Italy}
\address{$^4$ National Superconducting Cyclotron Laboratory and Joint Institute of Nuclear Astrophysics, Michigan State University, East Lansing, MI 48824, USA}
\ead{zhyx@ciae.ac.cn}

\begin{abstract}
Within the framework of ImQMD05, we study several isospin sensitive observables,  such as DR(n/p) ratios, isospin transport ratio (isospin diffusion), yield ratios for LCPs between the projectile region and mid-rapidity region for the reaction systems Ni+Ni, Zn+Zn, Sn+Sn at low-intermediate energies. Our results show that those observables are sensitive to the density dependence of symmetry energy, and also depend on the cluster formation mechanism. By comparing these calculations to the data, the information of the symmetry energy and reaction mechanism is obtained.
\end{abstract}

\section{Introduction}
The nuclear symmetry energy plays an important role in the properties of nuclei and neutron stars~\cite{Latti2001,Latti2004,Stein2005,Anna2006,Yakov2004}. To a good approximantion, it can be written as
\begin{equation}
E_{sym}=S(\rho)\delta^2.
\end{equation}
where $\delta=(\rho_n-\rho_p)/(\rho_n+\rho_p)$, is the isospin asymmetry; $\rho_{n}$, $\rho_{p}$, are the neutron, proton densities, and $S(\rho)$ describes the density dependence of the symmetry energy. Theoretical predictions for $S(\rho)$ from microscopic nucleon-nucleon interactions show large uncertainties, especially in the region of suprasaturation density \cite{Brown91,BALi08}. Constraining the density dependence of the symmetry energy has become one of the main goals in nuclear physics and has stimulated many theoretical and experimental studies \cite{BALi08,Danie02,Fuch06,Garg04,HSXu00,Tsang01,Shett04,Tsang04,LWCh04,qfli05,qfli06,TXLiu07,BALi05,Fami06,BALi97,BALi06,BALi00,BALi04,
Yong06,Tsang09,Gior10,Napo10, zhang05,zhang08}. Heavy Ion Collisions (HIC) with asymmetric nuclei provide a unique opportunity for laboratory studies of the density dependence of the symmetry energy because a large range of densities can be momentarily achieved during HICs. In theoretical studies with transport models, the isospin ratio observables which are constructed from the isospin contents of emitted nucleons or fragments, such as Y(n)/Y(p) and DR(n/p) for emitted nucleons\cite{ Fami06,BALi97, zhang08}, isospin transport ratios $R_i$ constructed from the isospin asymmetry of projectile residues (or emitted resource)\cite{Tsang04,LWCh04,Tsang09,zhang12}, and $R^{mid}_{yield}$, constructed from the yields of LCP between the mid-rapidity and projectile region\cite{Kohley}, have been prove to be primarily sensitive to the density dependence of the symmetry energy. By comparing the theoretical predictions to the experimental data, the sought-after constraints can be obtained.

One frequently utilized transport models to describe the heavy ion collisions is the Boltzmann-Uehling-Uhlenbeck (BUU) equation, which provides an approximate Wigner transform of the one-body density matrix as its solution\cite{Bertsch88}.
\begin{eqnarray}
\frac{\partial f}{\partial t}+v \cdot \nabla_{\mathbf{r}}f-\nabla_{\mathbf{r}}U \cdot \nabla_{\mathbf{p}}f && =
-\frac{1}{(2\pi)^6}\int d^3p_2d^3p_{2'}d\Omega\frac{d\sigma}{d\Omega}v_{12}\nonumber\\
&&\times \{ [ff_2(1-f_{1'})(1-f_{2'})]-[f_{1'}f_{2'}(1-f)(1-f_{2})]\nonumber\\
&&\times (2\pi)^3\delta^{3}(\mathbf{p}+\mathbf{p}_2-\mathbf{p}_{1'}-\mathbf{p}_{2'})\} \label{buueq}
\end{eqnarray}
The l.h.s. of this equation is the total differential of $f$ with respect to the time assuming
a potential $U$. Usually a Skyrme-parametrization of the real part of the G-matrix or Skyrme-like energy density functional are employed as the nucleonic potential which describe the influence of the different isospin asymmetric nuclear equation of state (asy-EOS). Stochastic extensions of these mean-field based approaches have been
introduced (see \cite{Chomaz04} and references therein). For instance,
in the so-called Stochastic Mean Field (SMF) \cite{Baran02} model fluctuations are
injected in coordinate space by agitating the spacial density profile. The r.h.s. of Eq. (\ref{buueq}) contains a Boltzmann collision integral which describes the influence of binary hard-core collisions and are realized by the test particles.

Another frequently utilized approaches, known as the Molecular Dynamics Model (QMD)that represent the individual nucleons as Gaussian "wave-packet" with mean values that move in according the Ehrenfest theorem; i.e. Hamilton's equations\cite{Aiche87}.
\begin{eqnarray}
\dot{\mathbf{r}_i}=\frac{\mathbf{p}_i}{m}+\nabla_{\mathbf{p}_i}\sum_{j}\langle V_{ij} \rangle=\nabla_{\mathbf{p}_i}\sum_{j}\langle H \rangle\\
\dot{\mathbf{p}_i}=-\nabla_{\mathbf{r}_i}\sum_{j}\langle V_{ij} \rangle=-\nabla_{\mathbf{r}_i}\sum_{j}\langle H \rangle
\end{eqnarray}
The expectation of the total Hamiltonian $<H>$ is obtained from the real part of the G-matrix or the Skyrme energy density functional, and it describes the influence of the different asy-EOS. The collision part in the QMD models are handled as same as the way in BUU type models but it is for nucleons rather than the test particles.

In this work, we choose to simulate nuclear collisions with the code ImQMD05 developed at the China Institute of Atomic Energy (CIAE), details of this code are described in Ref. \cite{zhang05,zhang08,zhang06,zhang07}, for studying several isospin ratio observables, such as DR(n/p), isospin transport ratios $R_i$, $R^{mid}_{yield}$ ratios for the yields of LCP between the mid-rapidity and projectile region and their relations to the fragmentation mechanism. For brevity, we limit our discussion here to the parameterization of the symmetry energy used in our calculations, which is of the form
\begin{equation}
S(\rho)=\frac{1}{3}\frac{\hbar^2}{2m}\rho^{2/3}_{0}(\frac{3\pi^2}{2}\frac{\rho}{\rho_{0}})^{2/3}+\frac{C_{s}}{2}(\frac{\rho}{\rho_{0}})^{\gamma_{i}}.\label{srho}\end{equation}
where $m$ is the nucleon mass and the symmetry coefficient $C_s=35.19MeV$. Using this particular parameterization, the symmetry energy at subsaturation densities increases with decreasing  $\gamma_i$, while the opposite is true for supranormal densities. In general, the EoS is labeled as stiff-asy for $\gamma_i>1$, and as soft-asy for $\gamma_i<1$. Finally, we also give a brief discussion the recent comparisons between the ImQMD05 and SMF calculations for further understanding the theoretical issue in the describing the reaction mechanism at low-intermediate energy heavy ion collisions.

\section{isospin ratio from nucleon to fragments}
Isospin ratios which are constructed from the isospin contents of fragments, such as $R(n/p)=Y(n)/Y(p)$ (or named as n/p ratio), DR(n/p) from neutron-rich and neutron-poor systems, isospin transport ratios $R_i$ and $R^{mid}_{yield}=2 Y_{LCP}$($y^0<0.5$)/$Y_{LCP}$(0.5$<y^0<$1.5), are sensitive to the density dependence of symmetry energy. In this section, we will check their sensitivities to the density dependence of symmetry energy and try to get the information of symmetry energy from them.
\subsection{n/p ratio and DR(n/p) ratio}
The neutron to proton ratio
$R_{n/p}=Y(n)/Y(p)$ of
pre-equilibrium emitted neutron over proton spectra was considered
as a sensitive observable to the density dependence of symmetry
energy\cite{BALi97}, because it has a straightforward link to the
symmetry energy. In order to reduce the sensitivity to uncertainties
in the neutron detection efficiencies and sensitivity to relative
uncertainties in energy calibrations of neutrons and protons, the
double ratio
\begin{equation}
DR(n/p)=R_{n/p}(A)/R_{n/p}(B)
\end{equation}
had been measured by Famiano and compared with the transport model
prediction\cite{Fami06,BALi97}.

\begin{figure}[htbp]
     \centering
 \includegraphics[angle=270,width=0.40\textwidth]{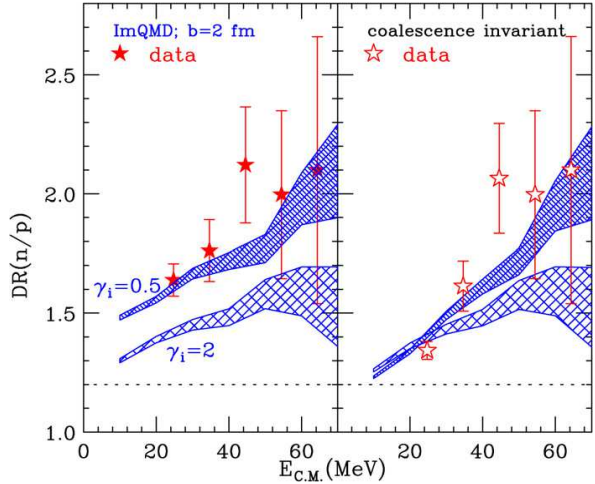}%
 \setlength{\abovecaptionskip}{40pt}
 \caption{\label{fig-dr}(Color online) (Left) DR(n/p) ratios for emitted free nucleons
and (Right) coalescent-invariant $DR(n/p)$ from the ImQMD simulations
are plotted as shadow region.}
  \setlength{\belowcaptionskip}{10pt}
\end{figure}

We performed calculations of collisions at an impact parameter
of $b=2 fm$ at an incident energy of $50 MeV$ per nucleon for two
systems: $A=^{124}Sn+^{124}Sn$ and
$B=^{112}Sn+^{112}Sn$ with ImQMD05 to study the
$DR(n/p)$ ratio for emitted nucleons\cite{zhang08}. The shaded regions in the left
panel of Fig.\ref{fig-dr} show the range, determined by uncertainties in
the simulations, of predicted double ratios $DR(n/p)=R_{n/p}(A)/
R_{n/p}(B)$ of the nucleons emitted between $70^{\circ}$  and
$110^{\circ}$  in the center of mass frame as a function of the
center of mass nucleon energy, for $\gamma_{i}=0.5$ and $2.0$. The
double ratios $DR(n/p)$ are higher for the EOS with the weaker
symmetry energy density dependence $\gamma_{i}=0.5$ than that for
$\gamma_{i}=2.0$ because the nucleons mainly emit from the lower
density region at intermediate energy HICs. Compare to the data on
$DR(n/p)$ for emitted nucleons(solid stars), the general trend of
data $DR(n/p)$ are qualitatively reproduced and the data seem to be
closer to the calculation employing the EOS with $\gamma_{i} =0.5$.
The right panel of fig.\ref{fig-dr} show the coalescence-invariant double ratio. The
coalescence-invariant double ratios are constructed by including all
neutrons and protons emitted at a given velocity, regardless of
whether they are emitted free or within a cluster. The data are
shown as open stars and the calculation results are shown as shaded
region. Here, the measurement and
simulation results illustrate that the fragments with $Z\ge2$
mainly contribute to the low energy spectra and do not affect the
high-energy $DR(n/p)$ data very much.

In order to constrain the range of $\gamma_{i}$ from the $DR(n/p)$
data that had been published, a series calculations for two systems,
$A=^{124}Sn+^{124}Sn$ and $B=^{112}Sn+^{112}Sn$, have been performed
by varying $\gamma_{i}=0.35, 0.5, 0.75, 1.0$ and $2.0$\cite{Tsang09}. Since the emitted nucleons
are mainly from the subnormal densities at this energies, the n/p
ratios of emitted nucleons are associated with the values of symmetry energy at subnormal density. Therefor, the $DR(n/p)$ ratio
should increase with decreasing $\gamma_{i}$. However, in the limit
of very small $\gamma_{i}\ll0.35$, the finite system completely
disintegrates and the $DR(n/p)$ ratio decrease and approaches the
limit of reaction system, $(N/Z)_{124}/(N/Z)_{112}=1.2$. As a
consequence of these two competing effects, the double ratio values
peak around the $\gamma_{i}=0.7$. Despite the large experiment
uncertainties for higher energy data, those comparisons definitely
rule out very soft ($\gamma_{i}=0.35$) and very stiff
($\gamma_{i}=2.0$) density dependence of symmetry energy. The $\chi^2$ analysis suggest that within a $2\sigma$ uncertainty, parameters of $\gamma_{i}$
fall in the range of $0.4\le\gamma_{i}\le1.05$ for the
$C_{s}=35.2MeV$.

\subsection{isospin transport ratio}
When the projectile and target nuclei come into contact, there can be exchange of nucleons between them. If the neutron to proton ratios of the projectile and target differ greatly, the net nucleon flux can cause a diffusion of the asymmetry $\delta$ reducing the difference between the asymmetries of two nuclei. This isospin diffusion process, which depends on the magnitude of the symmetry energy, affects the isospin asymmetry of the projectile and target residues in peripheral HICs. The isospin transport ratio $R_i$ has been introduced \cite{Tsang04} to quantify the isospin diffusion effects,
\begin{equation}
R_i=\frac{2X-X_{aa}-X_{bb}}{X_{aa}-X_{bb}}, \label{Ridef}
\end{equation}
where X is an isospin observable and the subscripts $a$ and $b$ represent the neutron rich and neutron-poor nuclei.  In this work, we use $a$ and $b$ to denote the projectile (first index) and target (second index) combination. where $\mathrm{a=^{124}Sn}$, and $\mathrm{b=^{112}Sn}$. We obtain the value of $R_{i}$ by comparing three reaction systems, $\mathrm{a+a}$, $\mathrm{b+b}$ and $\mathrm{a+b}$ (or $\mathrm{b+a}$). Construction of the transport ratio minimize the influence of other effects besides isospin diffusion effects on the fragment yields, such as preequilibrium emission and secondary decay, by rescaling the observable X for the asymmetric a+b system by its values for the neutron-rich and neutron-deficient symmetric systems, which do not experience isospin diffusion. Based on Eq. (\ref{Ridef}), one expects $R_i=\pm1$ in the absence of isospin diffusion and $R_i\sim0$ if isospin equilibrium is achieved. Eq. (\ref{Ridef}) also dictates that two different observables, $\mathrm{X}$, will give the same results if they are linearly related.
In one experiment, $\mathrm{X}$ was taken as the isoscaling parameter, $\alpha$, obtained from the yield of the light particles near the projectile rapidity\cite{Tsang01}, to measure the isospin diffusion ability in heavy ion collisions. In transport models \cite{Tsang04,LWCh04}, the isospin asymmetry $\delta$ of the projectile residues (emitting source) has been used to compute $R_i(\delta)$ because it is linearly related to the isoscaling parameters $\alpha$\cite{TXLiu07,Tsang01,Ono03}.

We analyze the amount of isospin diffusion with ImQMD05 by constructing a tracer from the isospin asymmetry of all emitted nucleons (N) and fragments (frag), including the heavy residue if it exists, with velocity cut $v^{N,frag}_z>0.5v^{c.m.}_{beam}$ (nearly identical results are obtained with higher velocity cut $v^{N,frag}_z> 0.7v^{c.m.}_{beam}$).  This represents the full projectile-like emitting source, and should be comparable to what has been measured in experiments.  Fig.2 shows the results of isospin transport ratios $R_i(X=\delta_{N,frag})$ (upright triangles) as a function of the impact parameter for a soft symmetry case ($\gamma_i=0.5$, open symbols) and a stiff symmetry case ($\gamma_i=2.0$, closed symbols). $R_i$ obtained with soft-symmetry case is smaller than those obtained with stiff-symmetry potential case. This is consistent with the expectation that higher symmetry energy at subnormal density leads to larger isospin diffusion effects (smaller $R_i$ values).
 \begin{figure}[htbp]
     \centering
 \includegraphics[angle=90,width=0.40\textwidth]{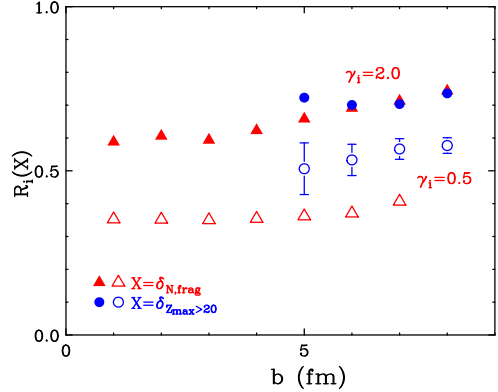}%
 \setlength{\abovecaptionskip}{20pt}
 \caption{(Color online) Isospin transport ratios as a function of impact parameter with two tracers for a soft symmetry case ($\gamma_i=0.5$, open symbols) and a stiff symmetry case ($\gamma_i=2.0$, closed symbols). Upright triangle symbols are for the tracer defined by the isospin asymmetry of all fragments and unbound nucleons with velocity cut ($v^{N,frag}_z>0.5v^{c.m.}_{beam}$), $X=\delta_{N,frag}$. Circles are for the tracer defined by the heaviest fragment with $Z_{max} > 20$ in projectile region, $X=\delta_{Z_{max} > 20}$.}
  \setlength{\belowcaptionskip}{0pt}
 \end{figure}

$R_i$ depends weakly on impact parameter over a range extending from central ($b=3fm$) to mid peripheral($b=8fm$) collisions. Interestingly, the isospin equilibrium and global thermal equilibrium are not reached even for central collisions. Our results show, that neither the effective interaction is sufficiently strong nor the collisions are sufficiently frequent (most of them are Pauli suppressed) to mix the projectile and target nucleons completely. These two effects prevent the combined system from attaining isospin equilibrium even in central collisions. With impact parameter increasing for $b>5fm$, the overlap region and thus the number of nucleons transferred from projectile and target decreases, causing the $R_i$ values to increase.

In peripheral collisions, most often, a large residue remains. If it decouples from the full emitting source before it equilibrates, it may experience a different amount of diffusion than the full emitting source examined by $X=\delta_{N,frag}$. To examine this, we constructed a tracer using the isospin asymmetry of the heaviest fragments with charge $Z_{max}>20$ in the projectile region. This tracer is mainly relevant to peripheral collisions as the central collisions are dominated by multifragmentation and very few large projectile fragments survive. The dependence of $R_i(X=\delta_{Z_{max}>20})$ for impact parameter $b\geq5fm$ is shown as open and closed circles in Fig. 2. The isospin transport ratios constructed from the different isospin tracers have different values especially in the case of $\gamma_i=0.5$. Stronger isospin equilibration (smaller $R_i$ values) is observed in the isospin transport ratios $R_i(X=\delta_{N,frag})$ constructed from nucleons and fragments than $R_i(X=\delta_{Z_{max}>20})$ constructed from the heaviest fragments with $Z_{max} > 20$. Since isospin diffusion mainly occurs through the low-density neck region,
and the system breaks up before isospin equilibrium, the asymmetry of the projectile and target residues do not achieve equilibrium and, larger $R_i(X=\delta_{Z_{max}\ge20})$ values result. In contrast, there is more mixing of nucleons from the target and projectile in the neck region due to the isospin diffusion. Consequently, rupture of the neutron-rich neck is predicted to result in the production of neutron-rich fragments at mid rapidity.

Since fragments are formed at all rapidities, we can examine the rapidity dependence of $R_i$ to obtain more information about the reaction dynamics. Fig. 3 shows $R_i$ as a function of the scaled rapidity $y/y_{beam}$. The symbols in the leftmost panel are experimental data obtained in Ref. \cite{TXLiu07} for three centrality gates. This transport ratio was generated using the isospin tracer $X=ln(Y(^7Li)/Y(^7Be))$ where $Y(^7Li)/Y(^7Be)$ is  the yield ratio of the mirror nuclei, $^7Li$ and $^7Be$ \cite{TXLiu07}. As expected the values of $R_i$ obtained from peripheral collisions (solid stars) are larger than those obtained in central collisions (open stars). For comparison, the ImQMD05 calculations of $R_i(X=\delta_{N,frag})$ are plotted as lines in the middle and right panels for a range of impact parameters. The middle panel contains the results from the soft symmetry potential ($\gamma_i=0.5$) while the right panel shows the results from the stiff symmetry potential ($\gamma_i=2.0$). The impact parameter trends and magnitude of the data are more similar to the results of the calculations from soft symmetry potentials ($\gamma_i=0.5$) for peripheral collisions.

We have performed the $\chi^2$ analysis for both
observables, $R_i$ and $R_i(y)$, for constraining the density dependence of symmetry energy. Using the
same 2$\sigma$ criterion, the analysis brackets the regions $0.45\le\gamma_i\le0.95$
is obtained. It is consistent with previous analysis on DR(n/p). However, the experimental trend of $R_i$ gated on the most central collisions (open stars) is not reproduced by the calculations. The experimental data indicate more equilibration for central collisions near mid rapidity while the transport model indicates more transparency. The equilibration in the E/A = 50MeV data may be the result of the impact parameter determination from charged particle multiplicity wherein the most central collisions are assumed to be the ones with highest charge particle multiplicity. For the most central events, a gate on the highest multiplicity, may select events in which more nucleon-nucleon collisions occur rather than a strict selection on the most central impact parameters.

 \begin{figure}[htbp]
     \centering
 \includegraphics[angle=90,width=0.45\textwidth] {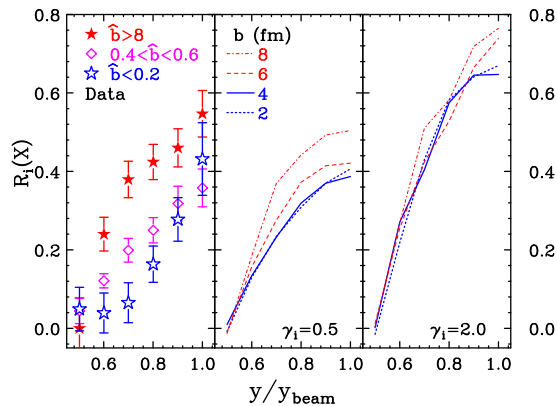} 
 \setlength{\abovecaptionskip}{20pt}
 \caption{(Color online) (Left panel) Experimental $R_i$ as a function of rapidity for three centrality gates [16]. (Middle panel) The calculated results of $R_i(X=\delta_{N,frag})$ as a function of rapidity for $b= 2, 4, 6, 8 fm$ for $\gamma_i=0.5$ and (Right panel) $\gamma_i=2.0$.}
  \setlength{\belowcaptionskip}{0pt}
 \end{figure}

\subsection{ $R^{mid}_{yield}$ ratios for light charged particles}
The yield ratios, $R^{mid}_{yield}$, for LCPs between the projectile region and mid-rapidity region, are defined as
\begin{equation}
R^{mid}_{yield}=\frac{2\cdot Yield(0.0\leq Y_r \leq0.5)}{Yield(0.5\leq Y_r \leq1.5)},
\end{equation}
where $Y_r=\frac{Y_{c.m.}}{Y^{c.m.}_{proj}}$ is the reduced rapidity.
It reflect the isospin migration ability and have been measured by \textit{Kohley.et.al.} for $\mathrm{^{70}Zn+^{70}Zn}$, $\mathrm{^{64}Zn+^{64}Zn}$, $\mathrm{^{64}Ni+^{64}Ni}$ at the beam energy of 35 MeV/nucleon for middle peripheral collisions \cite{Kohley}. The data show a clear preference for emission around the mid-rapidity region for more neutron-rich LCPs resulting from the isospin migration mechanism through the neck region between the projectile and target\cite{Baran04, zhang05}.
Theoretical study by SMF model\cite{Rizzo08} demonstrates that $R^{mid}_{yield}$ is sensitive to the density dependence of symmetry energy. The experimental trends are reproduced by SMF model. However, there are largest discrepancies on the reduced rapidity distribution for the yields of proton and $^3He$, and also on their values of $R^{mid}_{yield}$ as mentioned in Reference\cite{Kohley}. The discrepancies were thought to be related to the statistical decay of QP at later stages of the reaction \cite{Kohley}. In the points of reaction dynamics, the different fragmentation mechanism in the transport models simulations also lead to different behaviors of the rapidity distribution for LCPs besides the effects from secondary decay. Thus, it is instructive to study the rapidity distribution of LCP with ImQMD05.

\begin{figure}[htbp]
\centering
\includegraphics[angle=270,scale=0.45]{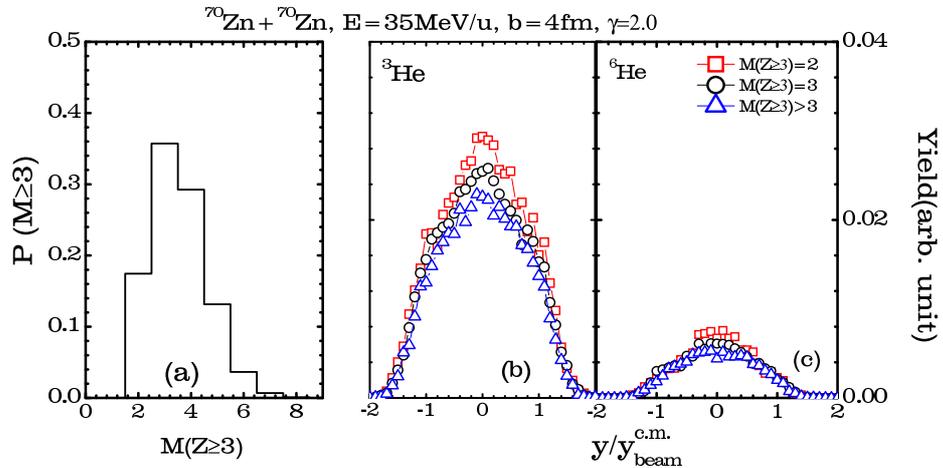}
\setlength{\abovecaptionskip}{45pt}
\caption{\label{ref-Fig.3}(Color online) (a) The multiplicity distribution for fragments with $Z\geq 3$ ($M(Z\geq 3)$). (b) is the reduced rapidity ($Y_r$) distribution for the yield of $^3He$ with binary (square symbols), ternary (circle symbols) and multi-fragmentation (triangle symbols) process. (c) is for $^6He$. All of those results are for $\mathrm{^{70}Zn+^{70}Zn}$ at E=35 MeV/u for b=4fm and $\gamma_i=2.0$.}
\setlength{\belowcaptionskip}{0pt}
\end{figure}

In Fig.\ref{ref-Fig.3}(a), we present the multiplicity distribution for fragments with $Z\geq 3$ for $\mathrm{^{70}Zn+^{70}Zn}$ at $E_{beam}$=35 MeV/nucleon and impact parameter b=4fm and $\gamma_i=2.0$. We find that \textbf{half} of events belongs to multi-fragmentation process which are defined by multiplicity for fragments with charge $Z>3$, i.e.,$M(Z\geq 3)>3$. The rest are the binary ($M(Z\geq 3)=2$) and ternary ($M(Z\geq 3)=3$) fragmentation events. It suggests that the binary, ternary fragmentation and multi-fragmentation coexist around 35MeV/nucleon. In Fig.\ref{ref-Fig.3} (b) and (c), we plot the reduced rapidity distribution for the yields of $^3He$ and $^6He$ obtained with three kinds of fragmentation process, binary (square symbols), ternary (circle symbols) and multi-fragmentation (triangle symbols) which are selected by $M(Z\geq 3)=2, 3$ and $>3$. The yields of $^3He$ and $^6He$ in Fig.\ref{ref-Fig.3} are normalized to per event. It is clear that the binary events produce more $^3He$ and $^6He$ at mid-rapidity relative to that produce in multi-fragmentation events. For $\gamma_i=2.0$ case, the yield of $^3He$ at $Y_r=0$ obtained with binary fragmentation events is 35\% larger than that with multi-fragmentation events. For neutron-rich LCP, for example, the yield of $^6He$ at $Y_r=0$ obtained in binary fragmentation events is 70\% larger than that in multi-fragmentation events due to isospin migration.

Fig.\ref{ref-Fig.4} shows the calculated results for the rapidity distribution of light charged particles \textit{p}, \textit{d}, \textit{t}, $^{3}He$, $^{4}He$ and $^{6}He$ for $^{64}Ni+^{64}Ni$ at b=4fm with 100,000 events.
The distributions are normalized with the yield at $Y_r=0$ for comparing with data in Ref \cite{Kohley} to understand the fragmentation mechanism, because the fragmentation mechanism mainly determine the shape of the rapidity distribution for LCP. The open circles are for $\gamma_i=0.5$ and solid symbols are for $\gamma_i=2.0$. The data are taken from the Ref \cite{Kohley} and plotted as stars. Both our calculations and data show that the width of distribution decreases with the mass of LCPs increasing. For the rapidity distributions of $^3H$ and $^3He$, the width of distribution for $^3H$ is smaller than that for $^3He$ due to the isospin migration. By comparing the simulated results to the data, the ImQMD05 calculations with stiffer symmetry energy can well reproduce the data at forward rapidity region ($Y_r>0$) for all \textit{p}, \textit{d}, \textit{t}, $^3He$, $^4He$ and $^6He$. For the backward rapidity region ($Y_r<0$), there are obvious differences between the results from ImQMD05 calculations and the data because the efficiency for detection of LCPs at the backward \cite{Kohley} are not included in this ImQMD05 calculations.

\begin{figure*}[htbp]
\centering
\includegraphics[angle=270,scale=0.4]{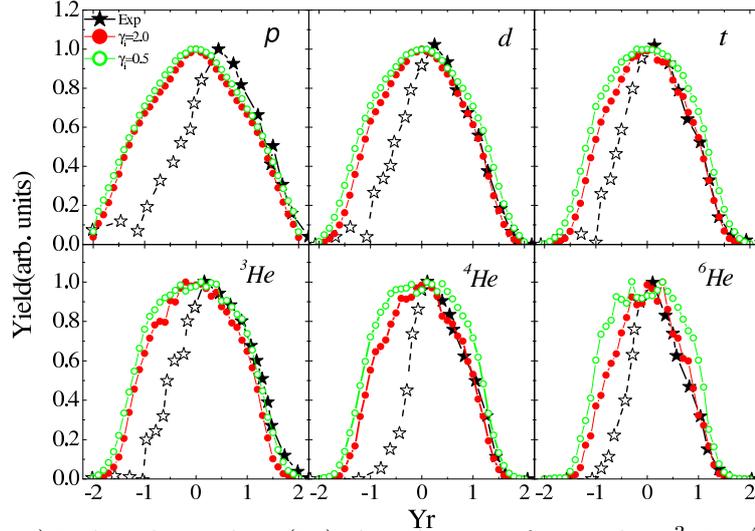}
\setlength{\abovecaptionskip}{35pt}
\caption{\label{ref-Fig.4}(Color online)Reduced rapidity ($Y_r$) distributions for \emph{p}, \emph{d}, \emph{t}, $^{3}He$, $^{4}He$ and $^{6}He$ fragments from the 35 MeV/nucleon $\mathrm{^{64}Ni+^{64}Ni}$  reaction for impact parameter b=4fm. The experimental data are shown as the stars. The ImQMD05 calculations for $\gamma_i=0.5$ are shown as the open circles and the solid circles are for $\gamma_i=2.0$. Each distribution is normalized with the yield at $Y_r=0$.}\label{ref-Fig.4}
\setlength{\belowcaptionskip}{0pt}
\end{figure*}

\begin{figure}[htbp]
\centering
\includegraphics[angle=270,scale=0.5]{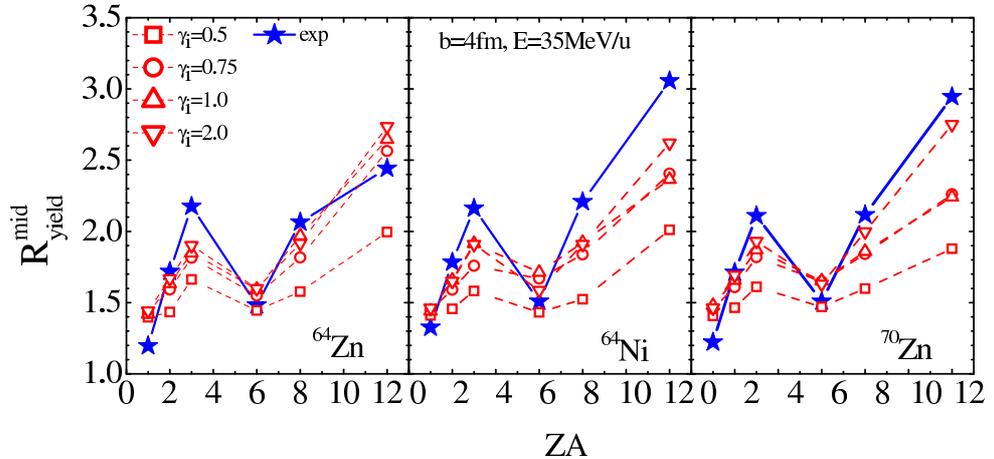}
\setlength{\abovecaptionskip}{45pt}
\caption{\label{ref-Fig.5}(Color online)$R_{yield}^{mid}$ values as a function of the charge times mass (ZA) for \textit{p} (ZA=1), \textit{d} (ZA=2),  \textit{t} (ZA=3), $^3He$ (ZA=6),  $^4He$ (ZA=8), $^6He$ (ZA=12). The open symbols are the results obtained with ImQMD05 for $\gamma_i=0.5, 0.75, 1.0$ and $2.0$. The solid stars are the data from \cite{Kohley}.}
\setlength{\belowcaptionskip}{0pt}
\end{figure}

In order to constrain the symmetry energy by the rapidity distribution of LCPs, we further calculate $R_{yield}^{mid}$ in a series of $\gamma_i=0.5, 0.75, 1.0, 2.0$. In Fig.\ref{ref-Fig.5}, we present the results of $R^{mid}_{yield}$ as a function of the AZ of emitted particles for three reaction systems $\mathrm{^{64}Zn+^{64}Zn}$, $\mathrm{^{64}Ni+^{64}Ni}$ and $\mathrm{^{70}Zn+^{70}Zn}$ at b=4fm. The open symbols are the results for $\gamma_i=0.5, 0.75, 1.0$ and $2.0$. The solid stars are the data from \cite{Kohley}. Since the isospin migration occurs in the neutron-rich neck region, the $R^{mid}_{yield}$ shows an increasing trend with the values of isospin asymmetry of LCPs increasing for the same element. The more neutron-rich the LCPs is, the larger the $R^{mid}_{yield}$ is. Furthermore, the calculated results show the values of $R^{mid}_{yield}$ for neutron-rich isotopes are sensitive to the density dependence of symmetry energy. The calculations with stiffer symmetry energy predict larger values of $R^{mid}_{yield}$ due to the stronger isospin migration effects. This conclusion is as same as the results obtained with SMF model \cite{Kohley}. As shown in Fig.\ref{ref-Fig.5}, the ImQMD05 calculations with stiffer symmetry energy ($\gamma_i\ge0.75$) can reasonably reproduce the data of $R^{mid}_{yield}$ as a function of AZ for $\mathrm{^{64}Zn+^{64}Zn}$. But our calculations underestimate the $R^{mid}_{yield}$ values of neutron-rich light charged particles, such as $^6He$, for the neutron-rich reaction systems $\mathrm{^{64}Ni+^{64}Ni}$ and $\mathrm{^{70}Zn+^{70}Zn}$. It could come from the lacking of fine structure effects of neutron-rich elements (such as neutron-skin, stability of lighter neutron-rich elements), and the impact parameter smearing effects in the transport model simulations. Even though our calculations can reproduce the $R^{mid}_{yield}$ data for $^{64}Zn+^{64}Zn$, the definitely constraints on the symmetry energy with the data of $R^{mid}_{yield}$ can not be obtained before we fix the problems on the theoretical predictions of $R^{mid}_{yield}$ for neutron-rich reaction system.

\subsection{remarks on the comparisons between ImQMD05 and SMF}
Up to date, there are several constraints on the symmetry energy with isospin sensitive observables, DR(n/p) raitos, isospin transport ratios $R_i$ and $R^{mid}_{yield}$ ratios, by adopting different type of transport models, such as QMD type and BUU type\cite{Tsang09,LWChen05,Rizzo08} . There are overlap between the results of symmetry energy from different approaches, but they are different in detail. Thus, further understanding the issues in the QMD type and BUU type would be crucial in theoretical studies for improving the constraints on the symmetry energy.

At the code level, both BUU and QMD models propagate particles classically under the influence of a mean field potential, which is calculated self-consistently the positions and momenta of the particles, and allow scattering by nucleon-nucleon collisions due to the residual interaction. The Pauli principle in both approaches is enforced by application of Pauli blocking factors. These similarities in implementation have lead to similarities in predictions for many collision observables \cite{Aich89}.

There are also significant differences in these approaches. In the BUU equations, each nucleon is represented by 200-1000 test particels that generate the mean field and suffer the collisions. In QMD, there is one test particle per nucleon. A-body correlations and cluster formation are not native to the original BUU approach; which is supposed to provide the Wigner transform of the one body density matrix. On the other hand, many-body correlations and fluctuations can arise from the A-body dynamics of QMD approach. Such A-body correlations are suppressed in BUU approach, but correlations can arise in both approaches from the amplification of mean field instabilities in spinodal region \cite{Chomaz04}. Collision algorithms in the QMD approach modify the momenta of individual nucleons, while in BUU approach, only the momenta of test particles are modified. Depending on the details of the in-medium cross sections that are implemented, the blocking of collisions can also be more restrictive for QMD than for BUU, leading to fewer collisions and therefore a greater transparency.

Fragments can be formed in QMD approaches due to the A-body correlations and these correlations are mapped onto the asymptotic final fragments by a minimum spanning tree algorithm. Serval different methods have been developed to allow BUU codes to calculate cluster production. In the Stochastic Mean Field (SMF) approach (one of the BUU type model), the time evolution of the one-body phase-space distribution $f$ is governed by the nuclear mean-field, two-body scattering, and a fluctuating (stochastic) term which causes the fragmentation \cite{Baran02,Baran04, Rizzo08,Colonna98}.
Since there are typically more than 100 test particles per nucleon, collision induced fluctuations are smaller in BUU than in QMD possibly suppressing the fragment formation rates.

\begin{figure}[htbp]
     \centering
 \includegraphics[angle=270,width=0.35\textwidth] {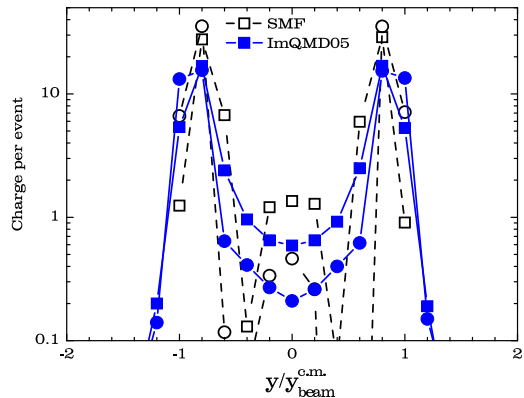} 
 \setlength{\abovecaptionskip}{30pt}
 \caption{\label{ref-zhangsmf}(Color online) The average charge number for $Z\ge3$ as a function of rapidity for $^{124}Sn+^{124}Sn$ at b=6,8fm with $\gamma_i=2.0$. The solid lines are the results from ImQMD05, and dashed lines are from SMF.}
  \setlength{\belowcaptionskip}{0pt}
\end{figure}

As an example, we present the results of average charge number for $Z\ge3$ as a function of rapidity obtained with ImQMD05 and SMF models for $^{124}Sn+^{124}Sn$ at $E_{beam}=50AMeV$ and b=6, 8fm in Fig.\ref{ref-zhangsmf}. The solid lines are the results from ImQMD05, and dashed lines are from SMF calculations\cite{zhmari}. Squares are for b=6fm, and circles are for b=8fm. The results obtained with QMD simulations show that the average charge number of $Z\ge3$ increase with rapidity increasing before $y/y^{c.m.}_{beam}\sim1$ in the forward region in the QMD simulations, and two peaks appear around the projectile and target rapidity region, respectively. 
In the SMF simulations, the peak appears at mid-rapidity besides two peaks around projectile and target rapidity. It clearly shown that the strict Pauli blocking in the QMD models simulations leads to a greater transparency than that in the SMF type simulations. The larger fluctuation in the QMD models lead to more fragments and light particles emitted than that from SMF predictions.
As results, the fragments distributed over the whole rapidity region and heavier fragment is, larger velocity is. 
In SMF calculations, the fragmentation process is less effective due to
the reduced amplitude of fluctuations and many-body correlations.
This enhances the appearance of binary and ternary processes in
semi-peripheral heavy ion reactions, according to the prominent role of
the mean-field dynamics. As a consequence, the intermediate mass fragments tend to distribute at mid-rapidity. It also lead that the the average charge number of $Z\ge3$ have a peak around mid-rapidity and narrower rapidity distribution of lighter clusters than the results from QMD models.

\section{Summary}
In summary, we have investigated the influences of the density dependence of the symmetry energy on several different isospin ratio observables, such as DR(n/p) ratio, isospin transport ratios $R_i$, the rapidity dependence of isospin transport ratio $R_i(y)$ and $R^{mid}_{yield}$ raitos (the yield ratios of LCP between the mid-rapidity and projectile region) with ImQMD05. The study shows that these isospin ratio observables are sensitive to the density dependence of symmetry energy. This conclusion is similar to conclusions reached using BUU approaches in the range of symmetry energies studied here. By comparing the calculated results to data, the very soft ($\gamma_i=0.35$) and very stiff symmetry energy ($\gamma_i=2.0$) are ruled out.

Cluster formation is important for intermediate energy heavy ion collisions, and it modifies the spectral double ratios at $E_{c.m.}<
40 MeV$. We also tested different tracers by constructing corresponding isospin transport ratios for them using different symmetry energies. For weakly density dependent symmetry energies (small $\gamma_i$) with large symmetry energies at sub-saturation densities, the values of $R_i$ for the residue tracer $X=\delta_{Z_{max}>20}$ are larger than those extracted from the entire emitting source, i.e., $X=\delta_{N,frag}$. The difference between these two tracers can be examined experimentally as a new probe of the symmetry energy and reaction mechanism.

By studying reaction systems $\mathrm{^{64}Zn+^{64}Zn}$, $\mathrm{^{64}Ni+^{64}Ni}$ and $\mathrm{^{70}Zn+^{70}Zn}$ at the beam energy of 35 MeV per nucleon and b=4fm within the framework of ImQMD05, we find that half of events belongs to the multi-fragmentation mechanism, and half of them is of binary and ternary fragmentation events. The binary and ternary events produce more light charged particles at middle rapidity, and the multi-fragmentation events broaden the reduced rapidity distribution for the yields of LCPs. Both the data and our calculations illustrate that the reaction systems seems more transparency and more fragments, light particles emitted. As results, the data of the reduced rapidity distribution for the yields of LCPs and $R_{yield}^{mid}$ as a function of AZ for $^{64}Ni+^{64}Ni$ can be well reproduced by the ImQMD05 calculations. For neutron rich reaction systems $\mathrm{^{64}Ni+^{64}Ni}$ and $\mathrm{^{70}Zn+^{70}Zn}$, our calculations underestimate the $R^{mid}_{yield}$ values of neutron-rich light charged particles, such as $^6He$, it could be cause by the lacking of fine structure effects for lighter elements in the transport models, and the impact parameter smearing effects.
\section{Acknowledgments}
This work has been supported by the Chinese National Science Foundation under Grants 11075215, 10979023, 10875031, 11005022,11005155, 10235030, and the national basic research program of China No. 2007CB209900. We wish to acknowledge the support of the National Science Foundation Grants No. PHY-0606007.


\section*{References}
\begin{thebibliography}{9}

\bibitem{Latti2001} J. Lattimer and M. Prakash, Ap. J. 550, 426 (2001).
\bibitem{Latti2004} J. Lattimer and M. Prakash, Science 304, 536 (2001).
\bibitem{Stein2005} A. Steiner et al., Phys. Rep. 411, 325 (2005).
\bibitem{Anna2006} A. L. Watts and T. E. Strohmayer, Ap. J. 637, L117
(2006).
\bibitem{Yakov2004} D. G. Yakovlev and C. J. Pethick, Annu. Rev. Astron. Astrophys.
42, 169 (2004).
\bibitem{Brown91} B. A. Brown, Phys. Rev. C 43, R1513 (1991).
\bibitem{BALi08} B. Li et al., Phys. Rep 464, 113 (2008).
\bibitem{Danie02} P. Danielewicz, R. Lacey, and W. G.Lynch, Science 298,
1592 (2006).
\bibitem{Fuch06} C. Fuch and H. Wolter, Eur. Phys. J. A 30, 5 (2006).
\bibitem{Garg04} U. Garg, Nucl. Phys. A 731, 3 (2004).
\bibitem{HSXu00} H. S. Xu et al., Phys. Rev. Lett. 85, 716 (2000).
\bibitem{Tsang01} M. B. Tsang et al., Phys. Rev. Lett. 86, 5023 (2001).
\bibitem{Shett04} D. V. Shetty et al., Phys. Rev. C 70, 011601 (2004).
\bibitem{Tsang04} M. B. Tsang et al., Phys. Rev. Lett.
92, 062701 (2004).
\bibitem{LWCh04} L.-W. Chen, C. M. Ko, and B.-A. Li, Phys. Rev. Lett.
94, 032701 (2005).
\bibitem{qfli05} Q. Li et al., Phys. Rev. C 72, 034613 (2005).
\bibitem{qfli06} Q. Li et al., Phys. Rev. C 73, 051601 (2006).
\bibitem{TXLiu07} T. X. Liu et al., Phys. Rev. C 76,
034603 (2007).
\bibitem{BALi05} B.-A. Li and L.-W. Chen, Phys. Rev. C 72, 064611
(2005).
\bibitem{Fami06} M. A. Famiano et al., Phys. Rev. Lett. 97, 052701 (2006).
\bibitem{BALi97} B. Li, C. Ko, and Z. Ren, Phys. Rev. Lett. 78, 1644
(1997).
\bibitem{BALi06} B. Li, L. W. Chen, G. C. Yong, and W. Zuo,
Phys. Lett. B 634, 378 (2006).
\bibitem{BALi00} B. Li, Phys. Rev. Lett. 85, 4221 (2000).
\bibitem{BALi04} B. Li, Nucl. Phys. A 734, 539c (2004).
\bibitem{Yong06} G. Yong, B. Li, and L. Chen, Phys. Rev. C 73, 034603
(2006).
\bibitem{Tsang09} M. B. Tsang, Y. Zhang, P. Danielewicz, M. Famiano, Z. Li,
W. G. Lynch, and A. W. Steiner, Phys. Rev. Lett. 102, 122701 (2009).
\bibitem{Gior10} B. Giordano et al., Phys. Rev. C 81, 044611 (2010).
\bibitem{Napo10} P. Napolitani et al., Phys. Rev. C. 81, 044619 (2010).
\bibitem{zhang05} Y. Zhang and Z. Li, Phys. Rev. C 71, 024604 (2005).
\bibitem{zhang08} Y. Zhang, P. Danielewicz, M. Famiano, Z. Li, W. G. Lynch,
and M. B. Tsang, Phys. Lett. B 664, 145 (2008).
\bibitem{zhang12} Yingxun Zhang, D.D.S.Coupland, P.Danielewicz, et.al., Phys.Rev.C85, 024602(2012).
\bibitem{Kohley} Z. Kohley, L.W. May, S. Wuenschel, et.al., Phys.Rev.C83, 044601(2011).

\bibitem{Bertsch88} G.F. Bertsch, S. Das Gupta, Phys.Rep. 4, 189(1988).
\bibitem{Chomaz04} P.Chomaz, M.Colonna, J.Randrup, Phys.Rep. 389, 263 (2004).
\bibitem{Baran02} V. Baran et. al, Nucl. Phys. A 703, 603 (2002).
\bibitem{Aiche87} J. Aichelin, A. Rosenhauer, G. Peilert, H. Stocker, and W. Greiner, Phys. Rev. Lett. 58, 1926 (1987).
\bibitem{zhang06} Y. Zhang and Z. Li, Phys. Rev. C 74, 014602 (2006).
\bibitem{zhang07} Y. Zhang, Z. Li, and P. Danielewicz, Phys. Rev. C 75,
034615 (2007).
\bibitem{Ono03} A. Ono, P. Danielewicz, W. A. Friedman, W. G. Lynch,
and M. B. Tsang, Phys. Rev. C 68, 051601(R) (2003).

\bibitem{Baran04} V. Baran et al, Nucl. Phys. A 730, 329 (2004)
\bibitem{Rizzo08} J. Rizzo et al, Nucl. Phys. A 806, 79 (2008).
\bibitem{zhmari} in priviate communication

\bibitem{LWChen05} L. W. Chen, C. M. Ko, B. A. Li, Phys. Rev. C 72, 064309 (2005).
\bibitem{Colonna98} M. Colonna et al., Nucl. Phys. A 642, 449 (1998).
\bibitem{Aich89} J.Aichelin, C.Hartnack, A. Bohnet, Zhuxia Li, G.Peilert, H.Stocker and W. Greiner, Phys.Lett. B 224, 34(1989).



\end{thebibliography}
\end{document}